\documentclass[reprint, amsmath, amssymb, aps, showkeys]{revtex4-2}
\usepackage{graphicx}
\usepackage{dcolumn}
\usepackage{bm}
\usepackage{amsmath}
\usepackage{float}
\usepackage{subfigure}
\usepackage{booktabs}
\usepackage{CJKutf8}
\usepackage{hyperref}
\usepackage{tikz,xcolor}

\hypersetup{
  colorlinks=true,
  linkcolor=blue,
  urlcolor=blue,
  citecolor=blue,
  linktoc=all
}
\definecolor{lime}{HTML}{A6CE39}
\DeclareRobustCommand{\orcidicon}{%
	\begin{tikzpicture}
	\draw[lime, fill=lime] (0,0) 
	circle [radius=0.16] 
	node[white] {{\fontfamily{qag}\selectfont \tiny ID}};
	\draw[white, fill=white] (-0.0625,0.095) 
	circle [radius=0.007];
	\end{tikzpicture}
	\hspace{-2mm}
}

\foreach \x in {A, ..., Z}{%
	\expandafter\xdef\csname orcid\x\endcsname{\noexpand\href{https://orcid.org/\csname orcidauthor\x\endcsname}{\noexpand\orcidicon}}
}

\begin{document}
\begin{CJK*}{UTF8}{gbsn}

\title{Constraining Kerr Supermassive Black Hole Properties Using Gravitational Waves from Inspiraling Stellar-mass Binary Black Holes}

\author{Jie Wu (吴洁)\orcidA{}$^{1,2}$ }
\author{Jin-Tao Yao (姚金涛)\orcidB{}$^{1,2}$ }
\author{Mengfei Sun (孙孟飞)\orcidC{}$^{1,2}$ }
\author{\\Jin Li (李瑾)\orcidD{}$^{1,2,3}$ }
\email{cqujinli1983@cqu.edu.cn}
\author{Zhoujian Cao (曹周键)\orcidE{}$^{4,5,6}$ }
\email{zjcao@bnu.edu.cn}

\affiliation{$^{1}$College of Physics, Chongqing University, Chongqing 401331, China}
\affiliation{$^{2}$Department of Physics and Chongqing Key Laboratory for Strongly Coupled Physics, Chongqing University, Chongqing 401331, China}
\affiliation{$^{3}$Institute of  Advanced Interdisciplinary Studies, Chongqing University, Chongqing 401331, China}
\affiliation{$^{4}$Department of Astronomy, Beijing Normal University, Beijing 100875, China}
\affiliation{$^{5}$Institute for Frontiers in Astronomy and Astrophysics, Beijing Normal University, Beijing 102206, China}
\affiliation{$^{6}$School of Fundamental Physics and Mathematical Sciences, Hangzhou Institute for Advanced Study, UCAS, Hangzhou 310024, China}

\begin{abstract}
We study the capability of future space-based gravitational-wave (GW) detectors to constrain supermassive black hole (SMBH) properties through observations of inspiraling stellar-mass binary black holes (BBHs) orbiting them.
Focusing on stable hierarchical triple systems, we model the BBH motion in Kerr spacetime and compute the modulated GW signals using the post Newtonian waveform combined with moving-source transformation.  
Based on the LISA configuration and second-generation time delay interferometry technology, we estimate parameter uncertainties with the Fisher information matrix.  
Our results show that the outer semimajor axis has the strongest influence on parameter precision, while the SMBH spin and eccentricity mainly affect their own uncertainties.  
For high-SNR signals, the SMBH mass and orbital parameters can be measured with relative uncertainties on the order of $10^{-5}$, while the spin magnitude and its orientation can be constrained to within a few percent.
Applying the method to M87*-like system, GW observations provide more precise measurements of the SMBH mass and spin compared with current electromagnetic observations, highlighting the potential of space-based GW astronomy to probe SMBH properties with high accuracy.
\end{abstract}

\maketitle
\end{CJK*}

\section{Introduction}
With the ongoing operation of the fourth observing run of ground-based gravitational-wave (GW) detectors, more than two hundred GW events have been confidently detected~\cite{GWTC4_1,GWTC4_2}.   
Most of these signals originate from binary black hole (BBH) mergers, providing valuable insights into the physics of strong-field gravity~\cite{GW250114,GW150914,Roadmap}.
However, due to the transient character of ground-based observations, probing the influence of the surrounding environment on BBH evolution remains challenging~\cite{environments,short_time}.  
Future space-based GW detectors, such as LISA~\cite{LISA}, Taiji~\cite{Taiji}, and TianQin~\cite{TianQin}, will open the low-frequency window and enable continuous monitoring of inspiraling BBHs over timescales of months to years~\cite{CaiRG2023}.  
Their long-duration observations will allow the accumulation of subtle dynamical imprints, providing a promising opportunity to investigate environmental effects on BBH systems~\cite{GW190521_LISA,environmental_effects}.

A representative case involves a stellar-mass BBH formed near a supermassive black hole (SMBH) in galactic nuclei, where the binary becomes gravitationally bound to the SMBH, forming a three-body system~\cite{Inayoshi:2017hgw}.  
Under the gravitational influence of the SMBH, the BBH orbits around it and experiences dynamical perturbations that induce characteristic modulations in the emitted GWs~\cite{Samsing:2020tda,Chen:2017xbi}.  
The leading-order effect is the Doppler shift arising from the orbital motion of the BBH around the SMBH~\cite{ChenYB_PRL}.  
At higher order, the SMBH exerts a long-term torque on the inner binary, causing de Sitter precession of its orbital angular momentum~\cite{de_Sitter_precession}.  
Systems with a large mutual inclination between the inner and outer orbits may also undergo Kozai–Lidov oscillations, where eccentricity and inclination exchange periodically over secular timescales~\cite{Relativistic_effects}.  
By tracing the cumulative signatures of such environmental effects over long observation periods, one can infer key parameters of the SMBH, such as its mass and spin~\cite{Randall:2018lnh,Cardoso:2021vjq}.

Current electromagnetic (EM) approaches for determining SMBH properties face several limitations.  
Dynamical mass estimates require resolving the SMBH’s sphere of influence, which is feasible only for a few nearby galaxies, limiting the sample size and increasing uncertainties at greater distances~\cite{Kormendy:2013dxa}.  
Many active galactic nuclei are heavily obscured by dust and gas, hindering reliable measurements of their central regions~\cite{Hickox:2018xjf}.  
Indirect techniques such as reverberation mapping and X-ray reflection spectroscopy depend on assumptions about geometry, inclination, and emissivity, introducing calibration errors and model-dependent biases in mass and spin estimates~\cite{Ho:2014pka,Reynolds:2020jwt}.  
Consequently, EM observations alone cannot yet provide precise and consistent measurements of SMBH mass and spin across different galactic environments.  
In this context, a BBH residing near an SMBH can serve as a gravitational probe, offering a complementary way to investigate SMBH properties through modulated GW signals beyond the reach of traditional EM observations.

For SMBH–BBH triple systems, the calculation of GW waveforms can generally be divided into two approaches.  
The first approach solves the full three-body dynamics numerically for systems in which the BBH orbits very close to the SMBH and cannot be approximated by a stable outer orbit.  
Such configurations are commonly referred to as binary extreme mass ratio inspirals (b-EMRIs)~\cite{b-EMRI}, where the waveform may contain EMRI-like features superposed on the BBH signal.  
This method applies to systems with small outer separations, accurately capturing strong three-body interactions but requiring substantial computational resources~\cite{b-EMRI1,b-EMRI2,b-EMRI3}.  
The second approach focuses on hierarchical triple systems and introduces phenomenological modifications to the waveform of an isolated BBH.  
When the orbital separation of the BBH around the SMBH is much larger than the inner separation of the binary, the system remains gravitationally bound and stable against tidal disruption, forming a long-lived configuration suitable for such treatment~\cite{three-body_systems}.  
In this framework, the effect of the outer orbital motion is represented by phase correction terms that describe slow modulations induced by the SMBH~\cite{ChenYB_PRL}.  
This approach is widely adopted for relatively stable systems and provides an efficient means to explore their parameter space.  
Effective field theory approaches have also been developed to describe hierarchical triple systems~\cite{Kuntz:2021ohi,Kuntz:2021hhm,Kuntz:2022onu}.  
In this work, we focus on short-timescale ($\sim$yr) dynamics in strongly hierarchical configurations, and therefore adopt a point-particle approximation, neglecting the backreaction of the BBH on the SMBH. 

Recent studies have investigated GWs from such hierarchical triple systems.  
Inayoshi~\textit{et al.} examined the phase drift in the inspiral waveform caused by the acceleration of the BBH center of mass, offering a way to constrain the contribution of nuclear formation channels to the observed BBH population~\cite{Inayoshi:2017hgw}.  
Randall and Xianyu showed that the orbital motion of a BBH around a tertiary mass introduces a time-dependent phase variation that can distinguish signals from different galactic environments, though their analysis neglected transverse motion and relativistic effects~\cite{Randall:2018lnh}.  
Fang~\textit{et al.} studied the influence of an SMBH’s spin on the evolution of a nearby BBH and found that the resulting spin-induced modulation of the waveform could be distinguishable~\cite{Fang:2019mui}.  
Yu and Chen demonstrated that when the de Sitter precession period of the inner binary is comparable to the observation time, the combined effects of precession and Doppler modulation allow the SMBH mass to be determined with percent-level precision~\cite{ChenYB_PRL}.  
Laeuger~\textit{et al.} and Yin~\textit{et al.} further investigated eccentric Schwarzschild orbits and circular equatorial Kerr orbits, respectively~\cite{Laeuger:2023qyz,Yin:2024nyz}.
Kuntz~\textit{et al.} showed that the transverse Doppler effect breaks the leading-order degeneracy between the SMBH mass and inclination, providing guidance for our analysis~\cite{transverse_Doppler_effect}. 
Some studies have also proposed using such systems to probe environmental effects, such as  GW190521~\cite{Sberna:2022qbn,GW190521_LISA}.  
These studies underscore the growing potential of GW observations to serve as precise probes of the dynamical environment around SMBHs and provide a theoretical foundation for further investigation into their measurable signatures.

Based on these previous studies, we further investigate the impact of the orbital motion of BBHs, particularly the frame-dragging effect induced by the spin of the SMBH, on the resulting GW waveforms and the prospects for constraining SMBH parameters with future space-based detectors.
In the orbital modeling, we consider general Kerr orbits that more accurately represent the spacetime around spinning SMBHs, allowing for non-equatorial and eccentric trajectories.  
For waveform construction, we adopt a high-order post-Newtonian (PN) inspiral model as the intrinsic BBH waveform and apply a moving-source transformation to compute the radiation from an orbiting BBH, including both transverse velocity components and relativistic corrections.  
For detection modeling, we employ the LISA configuration and implement the second-generation time-delay interferometry (TDI 2.0) response to simulate realistic space-based observations.  
Within this framework, we use the Fisher information matrix (FIM) method to quantify the precision with which SMBH parameters can be measured and assess the potential improvement achievable through complementary EM observations.  
This study focuses on waveform modulations induced purely by orbital motion and demonstrates that future LISA observations could provide meaningful constraints on the mass and spin of SMBHs through such GW signals.

The structure of this paper is organized as follows. 
In Sec.~\ref{sec:Triple_System}, we describe the configuration of the hierarchical triple system, including the adopted Kerr orbits and the simplifying assumptions applied in our analysis. 
Section~\ref{sec:GW_signal} reviews the PN waveform model, the moving-source transformation, and the TDI 2.0 response used to construct the GW signals.
In Sec.~\ref{sec:Methodology}, we present the FIM formalism employed for parameter estimation and discuss the selection of relevant parameters.
Section~\ref{sec:Results} reports the resulting constraints on SMBH parameters derived from the FIM analysis and includes a case study of the central black hole in M87.
Finally, Sec.~\ref{Conclusion} summarizes the main findings and outlines the key implications of this work.
Throughout this paper, we use geometrized units $c=G=1$, where $c$ is the speed of light and $G$ is the gravitational constant.

\section{Hierarchical Triple System}\label{sec:Triple_System}
\subsection{System Configuration and Coordinate Setup}\label{subsec:System_Configuration}
\begin{figure*}[ht]
    \begin{minipage}{\textwidth}
        \centering
        \includegraphics[width=0.95\textwidth,
        trim=0 0 0 0,clip]{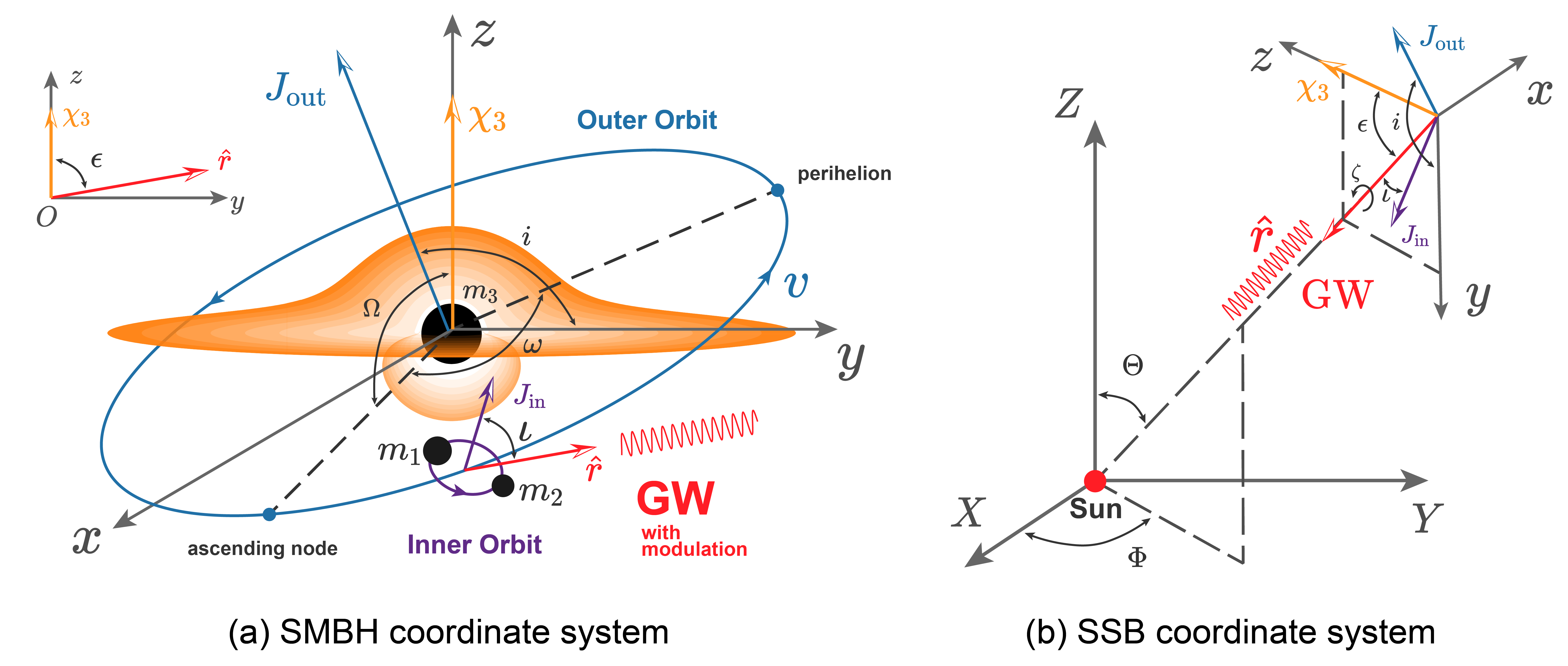}
        \caption{Schematic diagram of orbital elements and coordinate systems (not to scale). Panels (a) and (b) show the reference frames centered on the SMBH and on the solar system barycenter (SSB), respectively. The SMBH spin direction is chosen as the $z$-axis, and the $y$-axis is defined within the plane formed by the $z$-axis and the GW propagation direction $\hat{r}$, while the $x$-axis is given by $x = y \times z$. The angle $i$ denotes the inclination between the outer orbital angular momentum $J_{\mathrm{out}}$ and the $y$-axis. $\Omega$ is the position angle of the ascending node in the $xOz$ plane, and $\omega$ is the angle between the line of ascending node and the semimajor axis. The SMBH spin $\chi_3$ is tilted by an angle $\epsilon$ relative to $\hat{r}$, and the inclination $\iota$ is the inner orbital angular momentum $J_{\mathrm{in}}$ with respect to $\hat{r}$. The $yOz$ plane of the SMBH frame is rotated around $\hat{r}$ by an angle $\zeta$, and the sky location of the source in SSB coordinate system is $(\Phi, \Theta)$.}\label{fig:SMBH-BBH}
    \end{minipage}
\end{figure*}

We first introduce the SMBH–BBH system considered in this work.
The configuration of the triple system is illustrated in Fig.~\ref{fig:SMBH-BBH}, where the definitions of the coordinates and orbital elements follow those adopted in Ref.~\cite{Alexander:2005jz}.
The inner binary consists of the stellar-mass BBH with masses $m_1$ and $m_2$, while the outer binary is formed by the orbital motion around a SMBH of mass $m_3$.
Throughout this paper, subscripts \textit{in} and \textit{out} are used to distinguish quantities associated with the inner and outer orbits, respectively, and the indices 1 and 2 refer to the two components of the BBH, whereas the index 3 denotes the SMBH, as shown in Fig.~\ref{fig:SMBH-BBH}(a).
The relative orientation between the SMBH coordinate system and the SSB coordinate system is illustrated in Fig.~\ref{fig:SMBH-BBH}(b).
The BBH velocity components and SMBH spin components in the SMBH coordinate system can be transformed into those in the SSB coordinate system through a sequence of Euler rotations combined with a rotation about the propagation direction $\hat{r}$.
The detailed transformation procedure is provided in Appendix~\ref{App:Coordinate_transformation}.

When the mass of the SMBH satisfies $m_3 \gg  m_1, m_2$ and the outer semimajor axis $a_{\mathrm{out}}$ is much larger than the inner semimajor axis $a_{\mathrm{in}}$, the SMBH–BBH system behaves as a hierarchical triple configuration that remains dynamically stable over secular timescales.
In this regime, the inner BBH can be regarded as a compact subsystem whose center of mass follows a bound orbit around the SMBH, while the gravitational field of the SMBH acts as an external perturbation inducing slow modulations in the orbital elements of the inner binary.
Such a configuration provides a realistic description of stellar-mass BBHs residing in galactic nuclei, where dynamical interactions with the central SMBH can leave observable imprints on the emitted gravitational waves.

More specifically, we adopt a widely used stability criterion to assess whether the hierarchical triple systems considered in this work remain dynamically stable.
Following the empirical condition proposed by Eggleton and Kiseleva~\cite{Eggleton1995ApJ}, the dimensionless stability parameter is defined as $Y=[a_{\mathrm{out}}(1-e_{\mathrm{out}})]/[a_{\mathrm{in}}(1+e_{\mathrm{in}})]$.
The corresponding critical value is given by the fitting formula
\begin{equation}\label{eq:Yc}
  Y_c = 1 
+ \frac{3.7}{q_{\mathrm{out}}^{-1/3}} 
- \frac{2.2}{1 + q_{\mathrm{out}}^{-1/3}} 
+ \frac{1.4}{q_{\mathrm{in}}^{1/3}} 
  \frac{q_{\mathrm{out}}^{-1/3} - 1}{q_{\mathrm{out}}^{-1/3} + 1},
\end{equation}
where the inner mass ratio is defined as $q_{\mathrm{in}} = m_2/m_1 \le 1$, the outer mass ratio as $q_{\mathrm{out}} = m_3/m$, and the total mass of the BBH as $m = m_1 + m_2$.
According to this criterion, the triple system is \textit{stable} when $Y > Y_c$, whereas it becomes \textit{unstable} when $Y < Y_c$~\cite{Vynatheya2022MNRAS}.

\begin{figure}[ht]
    \begin{minipage}{\columnwidth}
        \centering
        \includegraphics[width=0.95\textwidth,
        trim=0 0 0 0,clip]{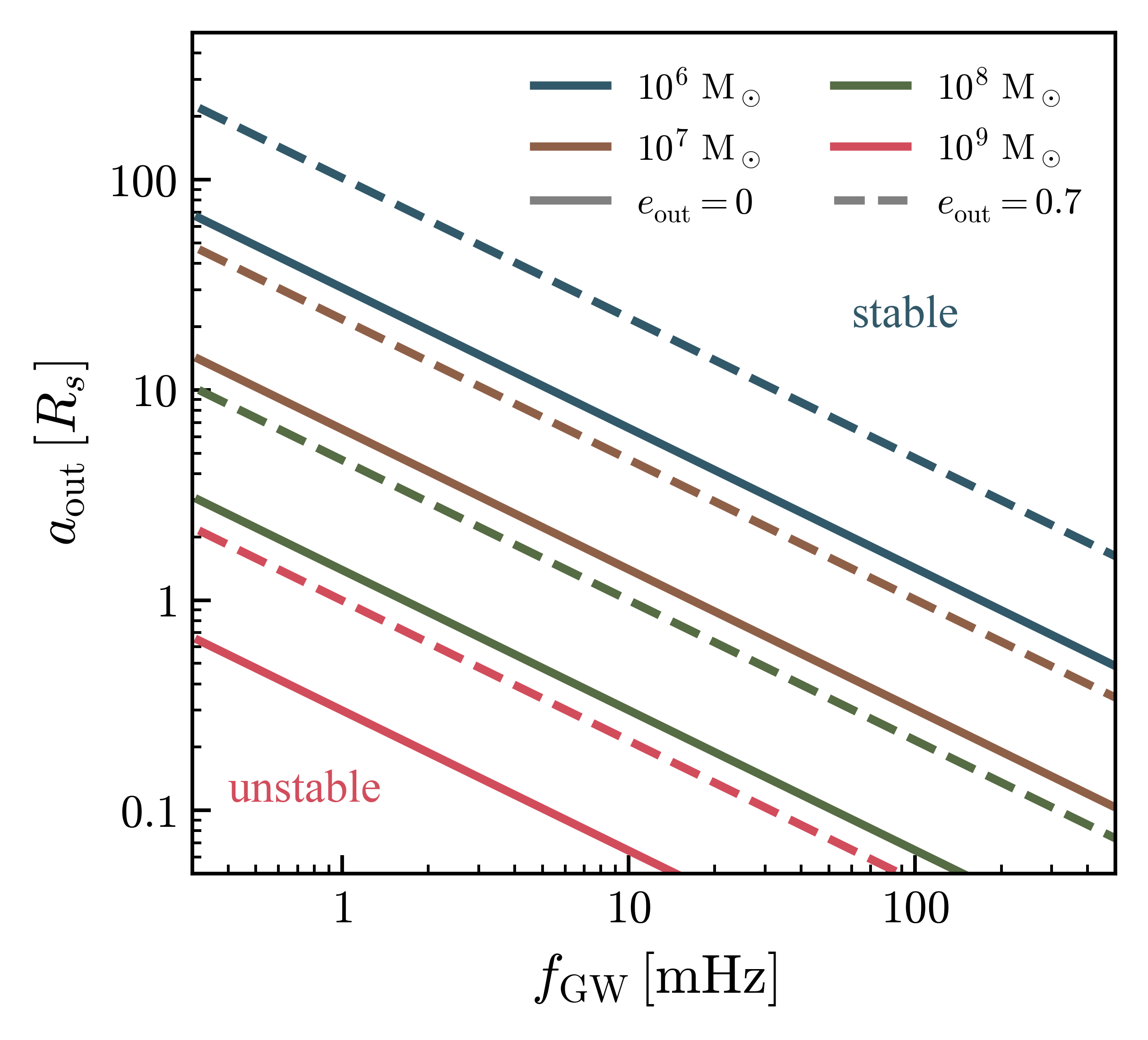}
        \caption{Variation of the outer semimajor axis $a_{\mathrm{out}}$ with the GW frequency $f_{\mathrm{GW}}$ of the inner BBH under the condition $Y = Y_c$. Both axes are shown on logarithmic scales. The solid line corresponds to an outer orbital eccentricity of $e_{\mathrm{out}} = 0$, while the dashed line corresponds to $e_{\mathrm{out}} = 0.7$.}\label{fig:stability}
    \end{minipage}
\end{figure}

To better characterize the stability boundary of the hierarchical triple systems considered in this work, we examine the condition $Y = Y_c$.
Our goal is to determine under what outer orbital configurations a BBH can remain dynamically stable.
According to Kepler’s law, the inner orbital frequency of the BBH is given by $\omega_{\mathrm{in}} = \sqrt{m / a_{\mathrm{in}}^{3}}$, and the corresponding GW frequency is $f_{\mathrm{GW}} = \omega_{\mathrm{in}} / \pi$.
We replace $a_{\mathrm{in}}$ in the stability criterion using $f_{\mathrm{GW}}$ to express the stability boundary in terms of an observable quantity.
Considering the large mass hierarchy between the SMBH and the BBH, we adopt the simplifying assumptions $e_{\mathrm{in}} = 0$, $q_{\mathrm{in}} = 1$, and $q_{\mathrm{out}} \gg 1$.
Under these conditions, Equation~(\ref{eq:Yc}) reduces to a simplified form 
\begin{equation}\label{eq:stability}
  a_{\mathrm{out} }[R_s]\simeq \frac{6.42}{1-e_{\mathrm{out} }} \left( \frac{10^7\mathrm{M_\odot } }{m_3}\times \frac{\mathrm{mHz}}{f_\mathrm{GW}}  \right)^{2/3} ,
\end{equation}
where $a_{\mathrm{out}}$ is expressed in units of the SMBH Schwarzschild radius, $R_s = 2m_{3}$.
For several representative SMBH masses, this dependence is illustrated in Fig.~\ref{fig:stability}.

From Eq.~(\ref{eq:stability}), the stability boundary, corresponding to the minimum value of the outer semimajor axis, scales proportionally with both the SMBH mass and the GW frequency raised to the power of $-2/3$.
Smaller values of $m_{3}$ or $f_{\mathrm{GW}}$ therefore require a larger $a_{\mathrm{out}}$ to maintain stability, and the same trend holds for higher outer orbital eccentricities $e_{\mathrm{out}}$. 
As shown in Fig.~\ref{fig:stability}, systems located below the curves are unstable, while those above are stable.
In this work, we consider $a_{\mathrm{out}}\sim10-800R_s$, with most cases being stable and a few requiring case-by-case evaluation.
In the parameter selection described in Sec.~\ref{subsec:Parameter_Selection}, we exclude configurations that do not satisfy the stability criterion.

\subsection{Orbital Motion in Kerr Spacetime}
\begin{figure}[ht]
    \begin{minipage}{\columnwidth}
        \centering
        \includegraphics[width=0.95\textwidth,
        trim=0 0 0 0,clip]{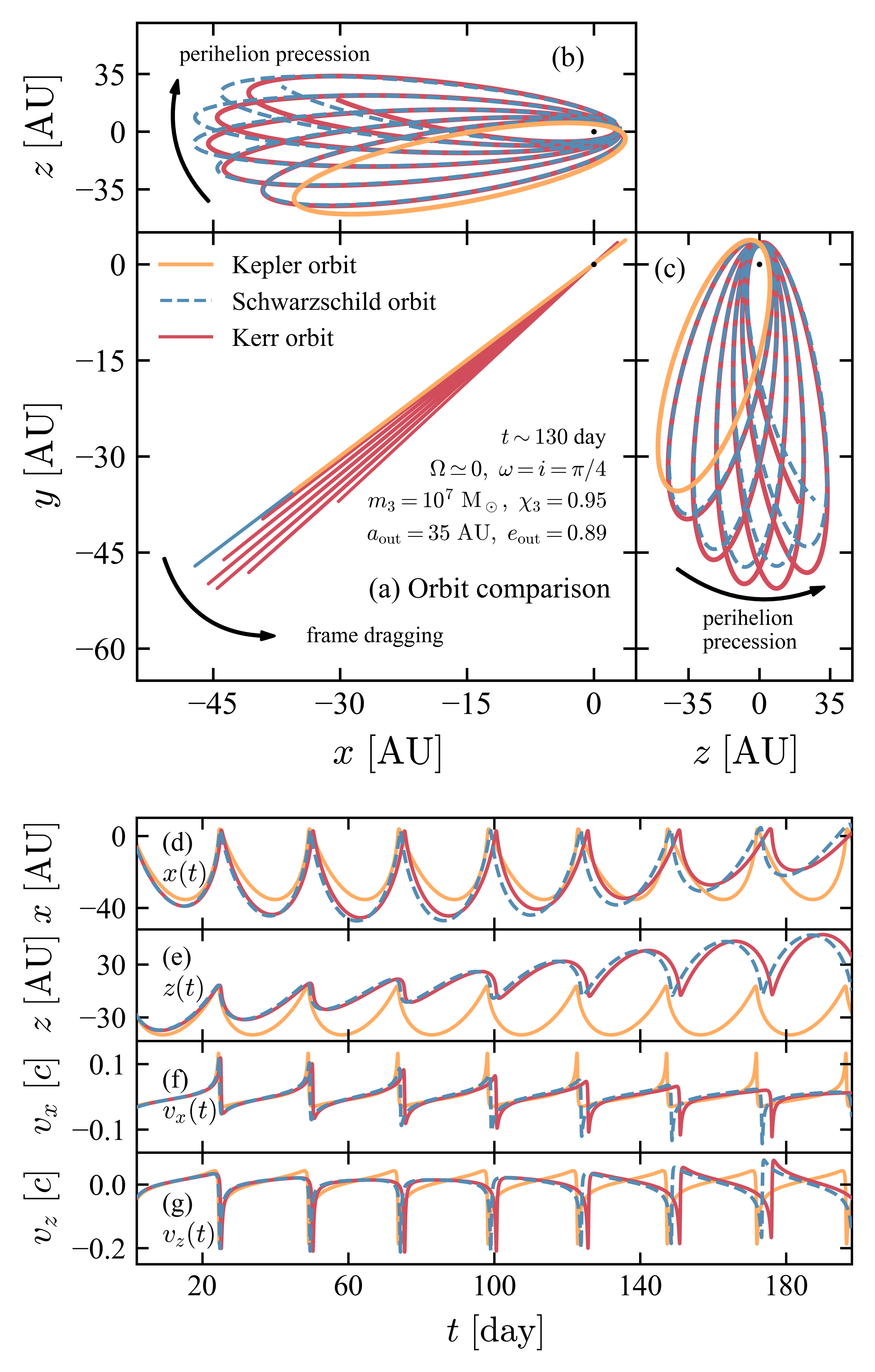}
        \caption{Comparison of the outer orbits around an SMBH in flat, Schwarzschild, and Kerr spacetimes. Panels (a)–(c) show the trajectories of the inner BBH from different viewing angles, with the specific orbital parameters indicated in panel (a). Panels (d) and (e) display the time evolution of the position components along the $x$ and $z$ directions, respectively. Panels (f) and (g) present the time evolution of the velocity components in the $x$ and $z$ directions.}\label{fig:kerr}
    \end{minipage}
\end{figure}

In modeling the orbital motion of the inner BBH around the SMBH, we adopt an outer orbit described in the Kerr spacetime. 
In Boyer-Lindquist coordinates, the metric takes the form~\cite{Kerr1963,Boyer1967,visser2008,Teukolsky2015}
\begin{align}
\mathrm{d}s^{2} = 
& -\left(1 - \frac{R_s r}{\Sigma}\right)\mathrm{d}t^{2}
+ \frac{\Sigma}{\Delta}\mathrm{d}r^{2}
+ \Sigma\,\mathrm{d}\theta^{2}  \nonumber \\[3pt]
& + \left(r^{2} + a^{2} + \frac{R_s r a^{2}}{\Sigma}\sin^{2}\theta\right)
   \sin^{2}\theta\,\mathrm{d}\phi^{2}  \nonumber \\[3pt]
& - \frac{2R_s r a}{\Sigma}\sin^{2}\theta\,
   \mathrm{d}t\,\mathrm{d}\phi ,
\end{align}
with
\begin{align}
\Sigma &= r^{2} + a^{2}\cos^{2}\theta , \\
\Delta &= r^{2} - R_s r + a^{2} , \\
a &= \chi_{3} m_{3} ,
\end{align}
where $\chi_{3}$ and $m_3$ are the dimensionless spin and mass of the SMBH.
Compared with the Newtonian Keplerian orbit and the Schwarzschild case, the Kerr orbit provides a more general and realistic description of motion around a spinning SMBH, naturally incorporating frame-dragging effects due to the SMBH spin.
Following the method described in Refs.~\cite{Yao1,Yao2}, we adopt a Cartesian coordinate system as defined in Fig.~\ref{fig:SMBH-BBH}(a) and numerically integrate the geodesic equations.
Specifically, we solve the second-order form of the geodesic equations with respect to the coordinate time $t$~\cite{LiXin}.
The geodesic equations can be expressed as~\cite{Carroll2019}
\begin{equation}
\frac{{\mathrm{d}}^{2} x^{\mu}}{{\mathrm{d}} t^{2}} 
+ \left( \Gamma^{\mu}_{\alpha \beta} 
- \Gamma^{0}_{\alpha \beta} \frac{{\mathrm{d}} x^{\mu}}{{\mathrm{d}} t} \right) 
\frac{{\mathrm{d}} x^{\alpha}}{{\mathrm{d}} t} \frac{{\mathrm{d}} x^{\beta}}{{\mathrm{d}} t} = 0 .
\end{equation}
We choose the perihelion as the initial position, and the calculation of the initial conditions only involves Kepler elements.
By substituting the initial conditions and numerically integrating the geodesic equations in the Kerr metric, the orbital trajectory can be obtained.
The detailed computational procedure follows Ref.~\cite{Yao1} and is not repeated here.
In this treatment, the BBH is approximated as a point particle following a geodesic in Kerr spacetime, neglecting internal structure effects on the outer motion. 
The validity of this approximation is discussed in Sec.~\ref{subsec:Assumptions_and_Validity}. 

In this way, we calculate a relatively extreme orbit to illustrate the differences among trajectories in different spacetimes.
The orbital plane is nearly perpendicular to the $xOy$ plane, as shown in Fig.~\ref{fig:kerr}, which clearly demonstrates the distinctions between various orbital configurations.
From Figs.~\ref{fig:kerr}(a)–(c), it can be seen that, in contrast to the static Keplerian orbit, both the Schwarzschild and Kerr orbits exhibit periapsis precession.
Furthermore, because the Keplerian and Schwarzschild spacetimes are spherically symmetric, their orbital planes remain fixed, while in the Kerr case, the frame-dragging effect caused by the SMBH spin leads to a precession of the orbital plane.
These features are also reflected in Figs.~\ref{fig:kerr}(d)–(g), which show the time evolution of each coordinate component.
Due to periapsis precession, the Schwarzschild and Kerr orbits quickly deviate from the Keplerian one, and over longer timescales, frame dragging further causes the Kerr orbit to diverge from the Schwarzschild orbit.

These differences in orbital motion lead to distinct velocities of the inner BBH, which in turn produce different modulations in the emitted GWs.
Such variations encode valuable information about the gravitational potential of the SMBH.
By analyzing these waveform modulations, we can not only measure the SMBH mass with high precision but also infer its spin magnitude, spin orientation, and the inclination of the orbital plane.
In particular, the characteristic signatures induced by frame dragging in Kerr spacetime provide a unique opportunity to probe the relativistic properties of the SMBH, offering insights beyond what can be achieved through EM observations.

\subsection{Assumptions and Validity}\label{subsec:Assumptions_and_Validity}
In this subsection, we clarify the assumptions adopted in our analysis and discuss their validity.
First, since the luminosity distance $D_{L}$ between the SMBH and the detector is much larger in scale than $a_{\mathrm{out}}$, we take $D_{L}$ as the distance when calculating the GWs emitted by the inner BBH.
Beyond ensuring that the selected parameters satisfy the stability criterion given by Eq.~(\ref{eq:Yc}) to maintain a stable hierarchical triple configuration, we assume a circular inner orbit with $e_{\mathrm{in}} = 0$.
This assumption is justified because $e_{\mathrm{in}}$ only affects the intrinsic GW waveform of the rest BBH and does not influence the modulation induced by the outer orbital motion~\cite{Laeuger:2023qyz}.
Since our primary objective is to investigate the constraints on the SMBH parameters, adopting a circular inner binary provides a reasonable and convenient simplification.

For a SMBH-BBH triple system, the emitted GWs can be divided into two components.  
The first originates from the inspiral of the inner BBH.  
In this work, we focus on GWs in the space-based detected frequency band ($\sim1-100~\mathrm{mHz}$), corresponding to the very early inspiral phase of the BBH, when the system is still tens to hundreds of years away from coalescence.  
This is different from many previous studies that consider the GW close to BBH coalescence observed by ground-based detectors.  
The second component arises from the motion of the BBH around the SMBH, which forms an EMRI system.  
These two kinds of GWs are emitted simultaneously and may overlap, but for the hierarchical triple configurations discussed in Sec.~\ref{subsec:System_Configuration}, such overlap can be safely neglected.  
A simple estimation gives the GW frequency of the outer binary as  
$\pi f_{\mathrm{out}}\simeq \sqrt{m_3/a^3_{\mathrm{out}}} $,
and substituting typical parameters yields  
\begin{equation}
f_{\mathrm{out}}[\mathrm{mHz}] \simeq 2.3\times10^{-3}
\left(\frac{10^{7}\,\mathrm{M_{\odot}}}{m_{3}}\right)
\left(\frac{100R_s}{a_{\mathrm{out}}}\right)^{3/2}.
\end{equation}
For convenience in order-of-magnitude estimates, we express $a_{\mathrm{out}}$ in units of $R_s = 2m_3$, which leads to an overall scaling $\propto 1/m_3$. The same convention applies to the expressions below. 
With this scaling, the corresponding frequency is about three orders of magnitude lower than that of the inner BBH and lies outside the sensitivity band of space-based detectors.  
Therefore, the GW signal is dominated by the emission from the inner BBH, and the contribution from the outer binary can be safely ignored.

In addition to the GWs emitted by the inner and outer binaries, several other effects arise due to the complexity of the triple system.  
From the perspective of phase modulation, the most significant contribution is the Doppler effect caused by the orbital motion of the inner BBH around the SMBH.  
This effect can be characterized by the annual variation of the Doppler phase shift, expressed as~\cite{ChenYB_PRL} 
\begin{equation}
\Omega_{\mathrm{Dp}}[\mathrm{rad/yr}] \simeq 2.2\times10^2
\left(\frac{10^{7}\,\mathrm{M_{\odot}}}{m_{3}}\right)
\left(\frac{100R_s}{a_{\mathrm{out}}}\right)^{3/2}.
\end{equation}  
The next leading-order effect is the de Sitter precession of the inner orbit, which can be written in a similar form as~\cite{ChenYB_PRL}  
\begin{equation}
\Omega_{\mathrm{dS}}[\mathrm{rad/yr}] \simeq \frac{1.7}{1-e^2_\mathrm{out} } 
\left(\frac{10^{7}\,\mathrm{M_{\odot}}}{m_{3}}\right)
\left(\frac{100R_s}{a_{\mathrm{out}}}\right)^{5/2}.
\end{equation} 
Compared with the de Sitter precession, the Doppler phase shift is much larger and has a stronger impact on the GW signal.
This is one of the reasons why we choose to neglect the de Sitter precession in our analysis.
Another reason is that the de Sitter precession affects the angular momentum of the inner BBH but does not originate from the variation of the outer orbit, which is the primary source of the moving-source effects we aim to study.
The influence of de Sitter precession on the estimation of SMBH parameters has already been discussed in previous works~\cite{ChenYB_PRL,Laeuger:2023qyz} and is not further addressed in this paper.
 
The frame-dragging effect induced by the SMBH spin evolves on longer timescales but can still influence the entire outer orbit, gradually altering its velocity components.  
Such variations make the system particularly sensitive to moving-source effects.
The orbital velocity of the inner BBH around the SMBH can be roughly estimated as
\begin{equation}
  v_{\mathrm{out}} \simeq 0.07
\left(\frac{100R_s}{a_{\mathrm{out}}}\right)^{1/2},
\end{equation}
which corresponds to the case of a circular orbit, while the perihelion velocity of an eccentric orbit can be even higher.  
At these velocities, noticeable relativistic effects are expected, especially over long observation periods.  
The influence of the outer orbit on the GW signal is therefore not limited to a simple phase correction but also includes relativistic effects arising from high-speed motion, all of which are taken into account in our moving-source transformation.

In our treatment of the outer orbit, the BBH is approximated as a point particle whose center-of-mass follows a geodesic in Kerr spacetime. 
The validity of this approximation can be assessed within the effective field theory framework developed in Refs.~\cite{Kuntz:2021ohi,Kuntz:2021hhm,Kuntz:2022onu}.
In this formalism, the gravitational multipole expansion is controlled by two small dimensionless parameters. 
In our setup, these correspond to the internal orbital velocity $v_{\rm in}$ and the hierarchical ratio $\varepsilon = a_{\rm in}/a_{\rm out}$, namely  
\begin{equation}
v_{\rm in} \simeq 1.2\times 10^{-2}
\left(\frac{m}{100\,\mathrm{M_{\odot}}}\times \frac{f_{\rm GW}}{1\,{\rm mHz}}\right)^{1/3},
\end{equation}
\begin{equation}
\begin{aligned}
\varepsilon \simeq\;& 3.7\times 10^{-4}
\left(\frac{m}{100\,\mathrm{M_{\odot}}}\right)^{1/3}
\left(\frac{10^7\,\mathrm{M_{\odot}}}{m_3}\right) \\
&\times
\left(\frac{1\,{\rm mHz}}{f_{\rm GW}}\right)^{2/3}
\left(\frac{100\,R_s}{a_{\rm out}}\right) .
\end{aligned}
\end{equation}
The effects of the BBH internal structure can be systematically expressed as powers of $v_{\rm in}$ and $\varepsilon$ (see Table I of Ref.~\cite{Kuntz:2021ohi}). For example, the contribution from angular momentum coupling is suppressed relative to the leading-order term by a factor of $\sim v_{\rm in}^2 \varepsilon^{3/2} \sim 10^{-9}$, which is negligibly small for the systems considered here. 
Higher-order corrections are even more strongly suppressed.  

Therefore, for the hierarchical configurations considered in this work, the influence of the BBH internal structure on the center-of-mass motion is negligible, and the geodesic approximation is well justified over the four-year observation timescale.  
Other effects, such as Lidov–Kozai oscillations, typically act on secular timescales much longer than the observation period, often of order $\gtrsim 10^3$ yr for the systems considered here~\cite{Prodan:2014dla,ChenYB_PRL}.  
As a result, the orbital configuration of the inner BBH evolves only weakly during the observation, and its inclination and eccentricity can be treated as approximately constant.  
In summary, our focus is on the overall influence of the outer orbit on the GW signal. 
The rest-frame waveform of the inner BBH is modeled as the inspiral waveform of an isolated BBH, while higher-order effects on the inner binary or processes operating on longer timescales are neglected.

\section{Gravitational-Wave Signal}\label{sec:GW_signal}
In general relativity, GWs contain two polarization modes, which can be expressed in tensor form as  
\begin{equation}
	h_{ij} = h_{+} e^{+}_{ij} + h_{\times} e^{\times}_{ij},
\end{equation}
where $h_{+}$ and $h_{\times}$ denote the plus and cross polarizations, and $e^{+}_{ij}$ and $e^{\times}_{ij}$ are the corresponding polarization tensors.  
These tensors can be constructed in the SSB coordinate system by defining a right-handed orthonormal basis~\cite{GWSpace,TDC}.  
For BBHs in the early inspiral phase, the PN approximation provides an accurate description of the GW waveform.  
We employ the 3.5 order PN waveform, which includes both linear spin-orbit (SO) and quadratic spin-spin (SS) coupling effects, referred to as the 3.5PN+SO+SS waveform~\cite{Wu_PN,Wu_NA}.  
The explicit analytic expressions for each PN order can be found in Refs.~\cite{PN_all,PN_SO,PN_SS}.
Considering that the stationary phase approximation may break down due to Doppler-induced anti-chirping effects~\cite{GW190521_LISA}, we construct the PN waveform entirely in the time domain. 
The frequency-domain signal used in FIM is obtained via a direct Fourier transform of the time-domain waveform. 

In four-dimensional spacetime, GWs exhibit the characteristics of boost weight zero and spin weight 2, which makes their transformation complicated.  
Representing GWs in a three-dimensional tensor form allows for a more straightforward and practical implementation of the transformation.  
Following our previous works~\cite{Wu_kick,Wu_GC}, we obtain the GW from a moving source by applying a three-dimensional Lorentz tensor transformation combined with a time-coordinate transformation.  
A complete and detailed derivation can be found in Ref.~\cite{moving}.  
For a source moving with an arbitrary velocity $\vec{v}$, the transformed GW can be written as  
\begin{equation}
	\begin{aligned}
		h_{ij}' & =h_{ij}+v^{k}h_{kl}v^{l}\frac{1}{(1-\hat{r}\cdot\vec{v})^{2}}[\hat{r}_{i}\hat{r}_{j} \\
		 & -\frac{\gamma}{1+\gamma}(\hat{r}_{i}v_{j}+v_{i}\hat{r}_{j})+\frac{\gamma^{2}}{(1+\gamma)^{2}}v_{i}v_{j}] \\
		 & +v^{k}h_{kj}\frac{1}{1-\hat{r}\cdot\vec{v}}[\hat{r}_{i}-\frac{\gamma}{1+\gamma}v_{i}] \\
		 & +v^{k}h_{ik}\frac{1}{1-\hat{r}\cdot\vec{v}}[\hat{r}_{j}-\frac{\gamma}{1+\gamma}v_{j}], 
		\end{aligned}
\end{equation}
with
\begin{equation}
\gamma = \frac{1}{\sqrt{1-v^{2}}}.
\end{equation}

In modeling the gravitational-wave signal from a moving source, we include both the effects of source motion and gravitational redshift in the time transformation.  
Let $t_S$ and $t_D$ denote the time coordinates in the source and detector frames, respectively.  
Their relation is given by the Lorentz transformation~\cite{moving}
\begin{equation}\label{eq:Lorentz_time}
    \frac{\mathrm d t_S}{\mathrm d t_D}
    =
    \frac{1}{\gamma\left(1-\vec{v}\cdot\hat{r}\right)}.
\end{equation}
This expression includes both longitudinal and transverse Doppler effects.
In a Kerr spacetime, the proper time $\tau_S$  satisfies $\mathrm d \tau_S^2 = - \mathrm d s^2$, which leads to~\cite{Carroll2019}
\begin{equation}\label{eq:Kerr_time}
\frac{\mathrm d \tau_S}{\mathrm d t_S}=
\sqrt{- \left(g_{tt}
+2g_{t\phi}\dot{\phi}
+g_{rr}\dot r^2
+g_{\theta\theta}\dot\theta^2
+g_{\phi\phi}\dot\phi^2
\right)}.
\end{equation}
Since Eq.~(\ref{eq:Lorentz_time}) already accounts for the transverse Doppler effect through the Lorentz factor, the spatial components in Eq.~(\ref{eq:Kerr_time}) may lead to double counting.  
To avoid this, we retain only the gravitational redshift associated with the time component $g_{tt}$, and write
\begin{equation}
    \frac{\mathrm d \tau_S}{\mathrm d t_S}
    = \sqrt{1 - \frac{R_s r}{\Sigma}} \, .
\end{equation}
In our setup, the BBH is located outside the ergosphere (i.e., $g_{tt}<0$), where this approximation is valid~\cite{Bardeen1972fi}.  
Since gravitational redshift and transverse Doppler effects are of the same order of magnitude, both contributions are consistently included~\cite{transverse_Doppler_effect}.
Combining the above relations, the total time transformation is
\begin{equation}
    \frac{\mathrm d \tau_S}{\mathrm d t_D}
    =\frac{\mathrm d \tau_S}{\mathrm d t_S}\cdot
    \frac{\mathrm d t_S}{\mathrm d t_D}.
\end{equation}
This provides a unified description valid for arbitrary source velocities in a strong gravitational field.

Furthermore, we consider the response of space-based detectors to GWs.  
The Michelson interferometer measures GWs through the relative change in the arm lengths.  
The fractional length variation can be obtained by contracting the GW tensor with the unit vectors along the arms, which yields the single-arm response in the time domain~\cite{GWSpace}.
To better reflect the configuration of future space-based GW detection, we adopt the second-generation TDI combination and compute the signals in the $X$, $Y$, and $Z$ channels following the method developed in Ref.~\cite{Wu_TDI}.

For space-based detectors, we focus on GWs in the millihertz frequency band and therefore adopt the configuration of LISA as a representative case.  
Similar missions such as Taiji share comparable orbital designs, but their different sensitivities may lead to quantitative variations in the final results~\cite{Wu_ppE}.  
Detectors operating in other frequency bands, such as DECIGO~\cite{DECIGO}, are beyond the scope of this study.  
Details of the LISA orbital configuration and its noise power spectral density (PSD) can be found in Refs.~\cite{LISA_orbit,LISA_PSD}.

In summary, the complete procedure for obtaining the GW signal is as follows (see Fig.~\ref{fig:process}).  
We first numerically solve the geodesic equations to determine the velocity components of the BBH in the SMBH coordinate system.  
These velocities are then transformed into the SSB coordinate system through a rotation transformation.  
Using the obtained velocity, we perform a Lorentz transformation on the stationary 3.5PN+SO+SS waveform to generate the waveform of the moving source.  
Finally, the GW signal is calculated through the LISA detector response, employing the TDI 2.0 to obtain the signals in the $X$, $Y$, and $Z$ channels.

\section{Methodology}\label{sec:Methodology}
\subsection{Data analysis}
\begin{figure*}[ht]
    \begin{minipage}{\textwidth}
        \centering
        \includegraphics[width=0.72\textwidth,
        trim=0 0 0 0,clip]{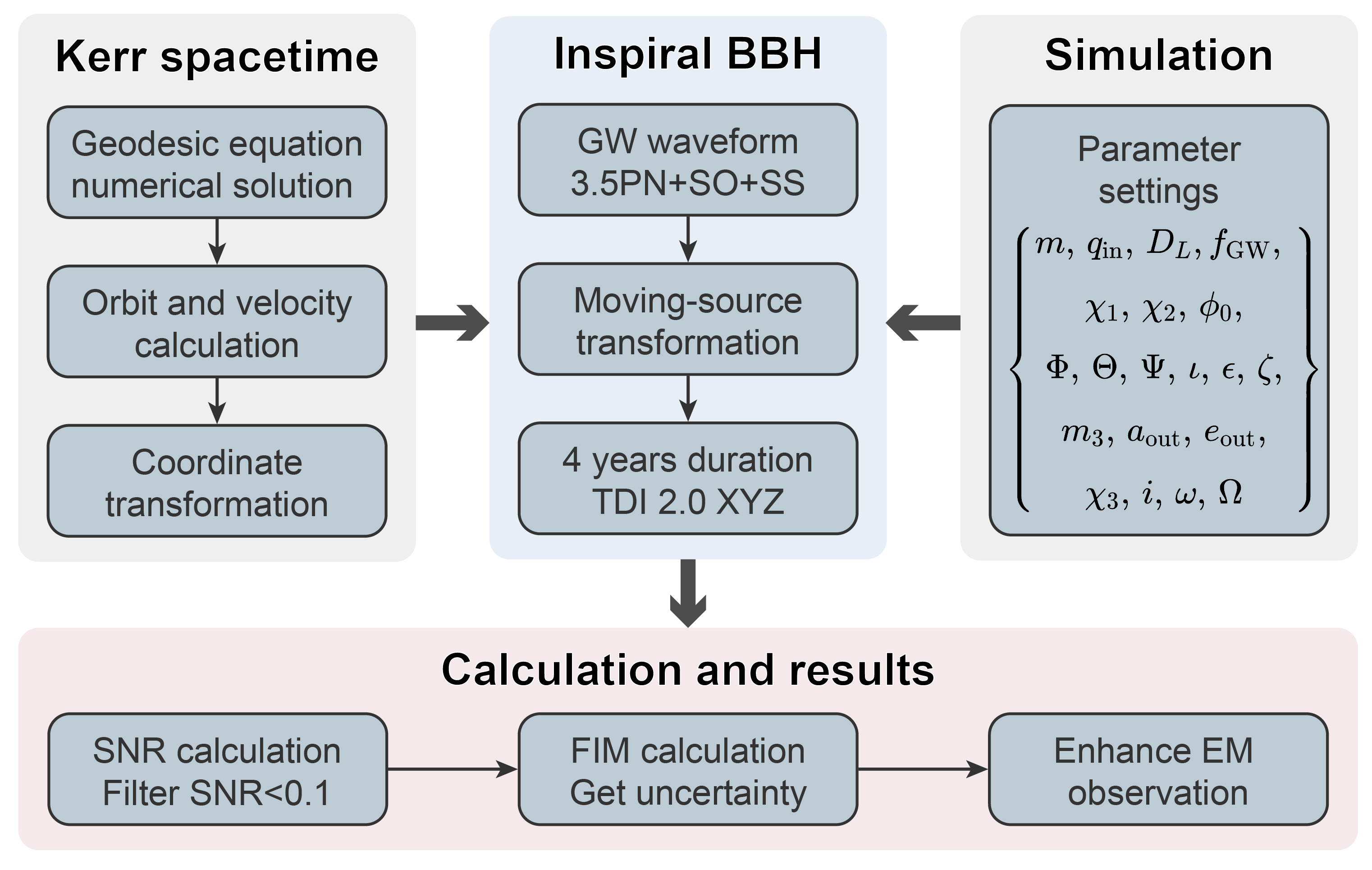}
        \caption{The process of simulating GW signals and calculating constraint results.}\label{fig:process}
    \end{minipage}
\end{figure*}
In GW data analysis, the inner product is usually defined as  
\begin{equation}
(a|b)=4\,\mathrm{Re}\!\left[\int_0^{\infty}\frac{\tilde{a}^*(f)\tilde{b}(f)}{S_n(f)}\,\mathrm{d}f\right],
\end{equation}
where $\tilde{a}(f)$ and $\tilde{b}(f)$ are the Fourier transforms of $a(t)$ and $b(t)$, and $S_n(f)$ is the detector PSD.  
For a time-domain GW signal $h(t)$, the signal-to-noise ratio (SNR) is defined as~\cite{SNR}
\begin{equation}
    \mathrm{SNR}^2 = (h|h).
\end{equation}
And the FIM is defined by  
\begin{equation}
    \Gamma_{ij} = \left(\frac{\partial h}{\partial\xi_i}\Bigg|\frac{\partial h}{\partial\xi_j}\right),
\end{equation}
where $\xi$ denotes the set of parameters.  
In our study, we use 20 parameters for the FIM calculation (see Fig.~\ref{fig:process}).
The inverse of the FIM, $\Sigma = \Gamma^{-1}$, corresponds to the covariance matrix, whose diagonal elements represent the parameter variances~\cite{FIM}.  
Accordingly, the 1-$\sigma$ uncertainty of a parameter $\xi_i$ is expressed as  
\begin{equation}
    \Delta\xi_i = \sqrt{\Sigma_{ii}}.
\end{equation}
In computing the FIM, we adopt the numerical differentiation approximation described in Refs.~\cite{FIM_ND,Wu_GB}, and the inverse matrix is evaluated following the method in Ref.~\cite{GWFAST} to estimate parameter uncertainties.  
For multiple channels, the combined SNR and FIM are obtained by summing the inner products over all channels~\cite{FIM_ND}.

The overall workflow of this study is illustrated in Fig.~\ref{fig:process}.
The GW signals we simulated in our research are all up to 4 years long, which is the same as the planned operating cycle of the space-based detectors.
After obtaining the GW signal, we first calculate the SNR and discard cases with $\mathrm{SNR}<0.1$.  
Although the detection threshold in real observations would be much higher, we adopt this lower cutoff to comprehensively explore the parameter space and evaluate the corresponding constraints.  
Subsequently, we compute the parameter uncertainties using the FIM, from which the constraints on the SMBH properties are derived.  
Finally, we analyze how these GW-based constraints can complement and improve the precision of future electromagnetic observations.
This procedure provides a consistent framework for quantifying the capability of space-based detectors to constrain SMBH parameters.

\subsection{Parameter Selection}\label{subsec:Parameter_Selection}
In this subsection, we describe the range of parameters adopted in our analysis, as summarized in Table~\ref{tab:parameters}.  
Here, $z$ denotes the redshift, and $\tau$ represents the time to coalescence.  
The parameters used directly or indirectly in the FIM calculation are derived from the values listed in this table.

\begin{table}[ht]
\centering
\renewcommand{\arraystretch}{1.5}
\caption{Parameter distribution used in this study~\cite{LISA_SOBBH}. $U[a,b]$ represents a uniform distribution from $a$ to $b$.}\label{tab:parameters}
\begin{tabular*}{\columnwidth}{@{\extracolsep{\fill}}cc@{}}
\hline
 Parameter & Distribution \\
\hline
$m$ [$\mathrm{M_\odot } $]
& $U[20,160]$ \\
$q_\mathrm{in}$ 
& $U[0.3,1]$\\
$\log_{10}(z)$ 
& $U[-3,-1]$\\
$\tau$  [yrs]
& $U[10,1000]/(1+z)$\\
$\chi_1,\chi_2$ 
& $U[-0.9,0.9]$\\
$\phi_0,\Phi,\Psi,\zeta,\omega,\Omega$ [rad]
& $U[0,2\pi]$ \\
$\iota,\Theta,\epsilon,i$ [rad]
& $\arccos(U[-1,1])$ \\
$m_3$ [$\mathrm{M_\odot } $]
& $[10^6,10^7,10^8,10^9]$ \\
$\chi_3$ 
& $U[0,0.95]$\\
$\log_{10}(a_\mathrm{out}\,[R_s])$ 
& $U[1,3]$\\
$e_\mathrm{out}$ 
& $U[0,0.9]$\\
\hline
\end{tabular*}
\end{table}

\begin{figure*}[ht]
    \begin{minipage}{\textwidth}
        \centering
        \includegraphics[width=0.95\textwidth,
        trim=0 0 0 0,clip]{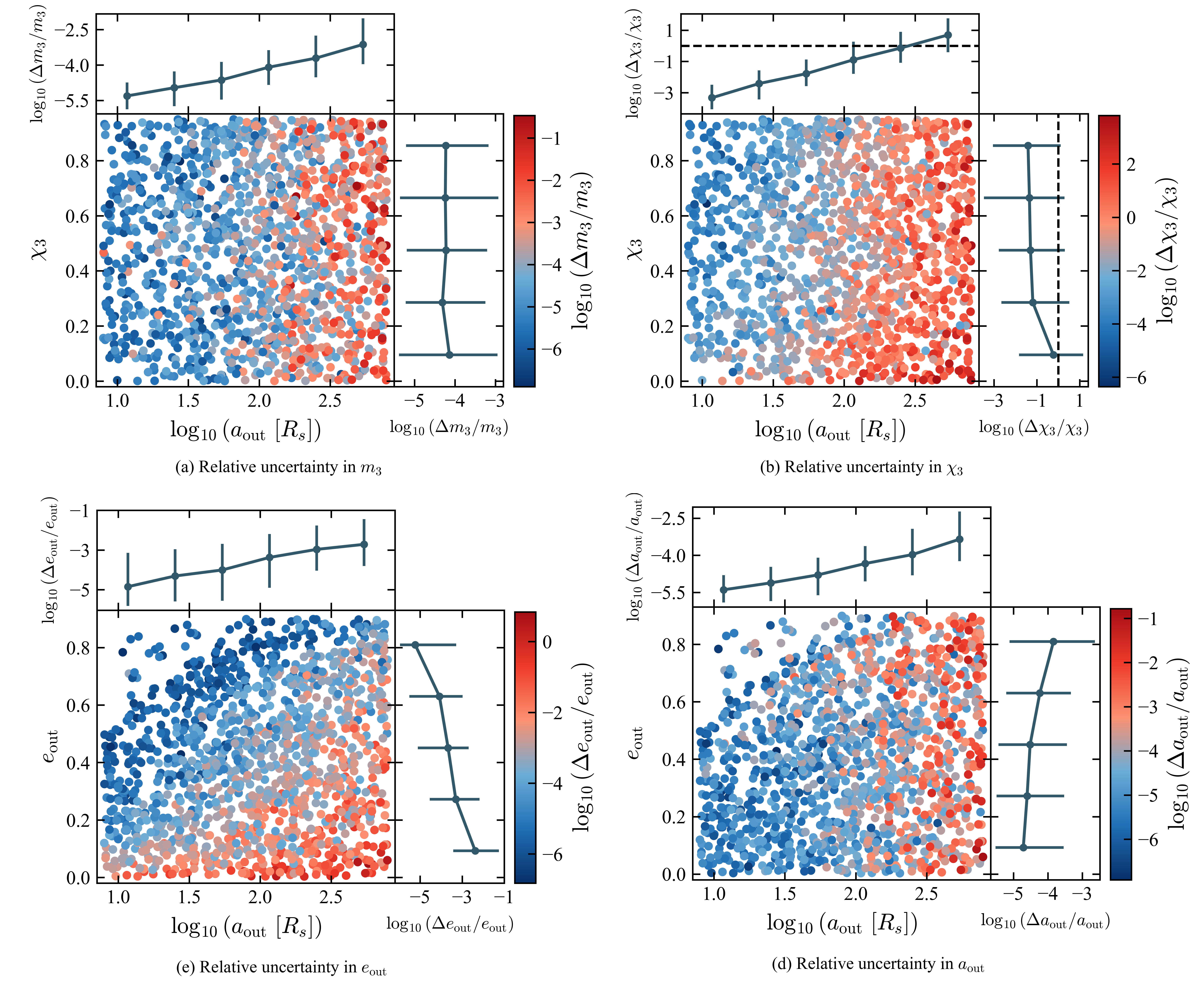}
        \caption{Distributions of the relative uncertainties of SMBH parameters with respect to the outer orbital parameters. Panels (a) and (b) show the relative uncertainties in $m_{3}$ and $\chi_{3}$ for systems with varying $a_{\mathrm{out}}$ and $\chi_{3}$, while panels (c) and (d) present the relative uncertainties in $e_{\mathrm{out}}$ and $q_{\mathrm{out}}$ for systems with varying $a_{\mathrm{out}}$ and $e_{\mathrm{out}}$. Each point corresponds to a simulated hierarchical triple system, and its color represents the relative parameter uncertainty. The marginal panels along the top and right sides display the median values and standard deviations of the one-dimensional distributions for each parameter. The black dashed line represents a relative uncertainty of 1}\label{fig:relative_uncertainty}
    \end{minipage}
\end{figure*}

From $z$ and $\tau$, we can calculate the luminosity distance $D_L$ and the initial GW frequency $f_{\mathrm{GW}}$ as~\cite{Wu_PN}  
\begin{equation}
    f_{\mathrm{GW}} = \frac{1}{8\pi}
    \left(\frac{1}{5}\,\mathcal{M}_{c}^{5/3}\,\tau\right)^{-3/8},
\end{equation}
\begin{equation}
    D_{L} = \frac{(1+z)}{H_{0}}
    \int_{0}^{z} 
    \frac{\mathrm{d}z'}{\sqrt{\Omega_{m}(1+z')^{3}+\Omega_{\Lambda}}},
\end{equation}
where $\mathcal{M}_{c} = (m_{1}m_{2})^{3/5}/(m_{1}+m_{2})^{1/5}$ is the chirp mass.  
We adopt the $\Lambda$CDM cosmological model with parameters from the \textit{Planck 2018 results}~\cite{Planck_2018}:  
the Hubble constant $H_{0}=67.37~\mathrm{km\,s^{-1}\,Mpc^{-1}}$,  
the matter density parameter $\Omega_{m}=0.315$,  
and the dark energy density parameter $\Omega_{\Lambda}=0.685$.

Following the procedure described above, we obtain the distributions of all relevant parameters and generate a dataset using weighted sampling.  
We consider four representative SMBH masses and construct in total more than two thousand hierarchical triple systems that satisfy the stability criterion.  
These configurations form the basis for evaluating the GW observables and parameter estimation in the following analysis.

\section{Results}\label{sec:Results}
\subsection{Constraints on SMBH Parameters}
With the simulated dataset established, we now investigate how LISA observations of stellar-mass BBHs orbiting SMBHs can constrain the properties of the central black hole.  
Using the FIM approach introduced earlier, we calculate the expected uncertainties of the SMBH parameters for different system configurations and examine how these constraints depend on the orbital and source parameters.

We first examine how the outer orbital parameters of the hierarchical triple system affect the parameter estimation results, as shown in Fig.~\ref{fig:relative_uncertainty}.  
Different outer orbital configurations lead to different modulations in the GW signal, resulting in variations in the parameter uncertainties.  
In some cases, clear trends appear between the parameter values and their corresponding uncertainties.  
In particular, $a_{\mathrm{out}}$ has the most significant impact on all results, with smaller values of $a_{\mathrm{out}}$ leading to smaller relative uncertainties.  
This is because a smaller $a_{\mathrm{out}}$ corresponds to a shorter orbital period, which causes stronger phase modulation in the GW signal, and faster orbital motion produces more pronounced relativistic effects, improving the precision of parameter constraints.

From Figs.~\ref{fig:relative_uncertainty}(a) and (b), it can be seen that the value of $\chi_{3}$ has only a minor effect on the parameter uncertainties.  
For the relative uncertainty of $m_{3}$, variations in $\chi_{3}$ hardly cause any noticeable change.  
In contrast, $\chi_{3}$ has some influence on its own relative uncertainty, especially when $\chi_{3}$ is small.  
In this regime, the frame-dragging effect produced by the SMBH spin is weaker, leading to less precise constraints on $\chi_{3}$.  
When $\chi_{3}$ becomes larger, particularly for $\chi_{3}>0.5$, the difference in results is negligible.  
Overall, compared with the influence of $a_{\mathrm{out}}$, changes in $\chi_{3}$ have only a limited impact on the relative parameter uncertainties.

From Figs.~\ref{fig:relative_uncertainty}(c) and (d), it is clear that variations in $e_{\mathrm{out}}$ have a significant impact on the uncertainty of its own parameter estimation.  
This trend is straightforward to understand, since a larger $e_{\mathrm{out}}$ leads to stronger modulation of the outer orbital motion and results in a smaller relative uncertainty of $e_{\mathrm{out}}$ itself.  
In contrast, the influence of $e_{\mathrm{out}}$ on the relative uncertainty of $a_{\mathrm{out}}$ shows the opposite behavior.  
Smaller values of $e_{\mathrm{out}}$ correspond to tighter constraints on $a_{\mathrm{out}}$.  
We suspect that this pattern may arise from the stability criterion applied to the system selection.  
As indicated by the missing region in the upper-left corner of the figure, systems with large $e_{\mathrm{out}}$ and small $a_{\mathrm{out}}$ do not satisfy the stability condition.  
Therefore, we infer that $e_{\mathrm{out}}$ itself has only a weak direct influence on the uncertainty of $a_{\mathrm{out}}$.  
The trend observed in Fig.~\ref{fig:relative_uncertainty}(d) likely results from the statistical correlation introduced by the stability criterion, under which systems with larger $e_{\mathrm{out}}$ tend to have larger $a_{\mathrm{out}}$, leading to higher relative uncertainties in $a_{\mathrm{out}}$.

\begin{figure}[ht]
    \begin{minipage}{\columnwidth}
        \centering
        \includegraphics[width=0.95\textwidth,
        trim=0 0 0 0,clip]{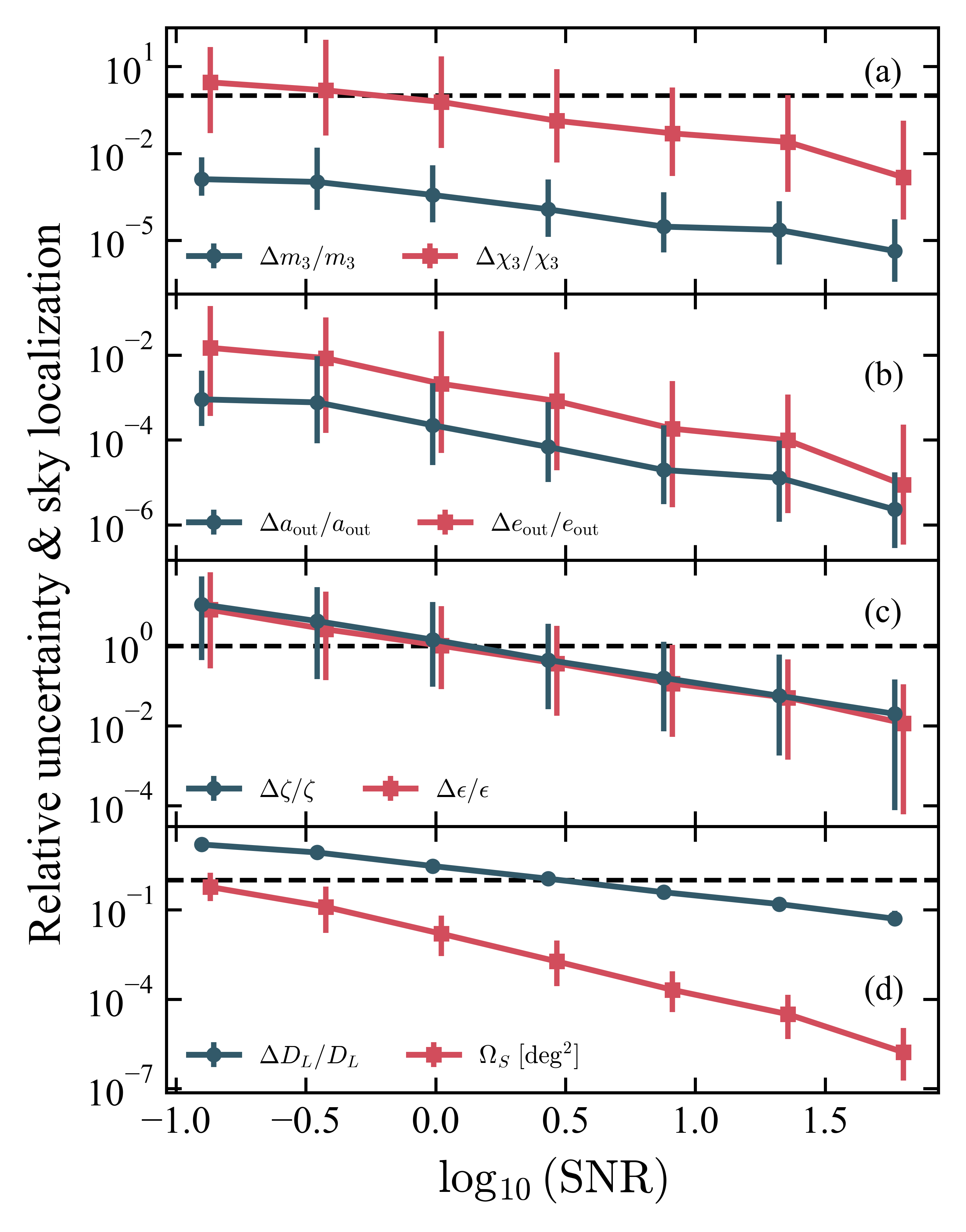}
        \caption{Variation of the relative parameter uncertainties and sky localization with SNR. Panels (a)–(d) show the results for the intrinsic SMBH parameters, outer orbit parameters, angular parameters, and sky localization, respectively. Each point represents the median value, and the error bars indicate the 16th and 84th percentiles. The black dashed line marks the level where the relative uncertainty equals 1.}\label{fig:SNR_error}
    \end{minipage}
\end{figure}

In summary, among the SMBH-related parameters, $a_{\mathrm{out}}$ has the strongest influence and affects the constraint results of all parameters, while $\chi_{3}$ and $e_{\mathrm{out}}$ mainly impact their own relative uncertainties.  
We next examine the effects of other parameters.  
For simplicity, we use the SNR as an indicator, since it also reflects the detectability of the GW signal.  
The dependence of the relative uncertainties of the SMBH parameters on the SNR is shown in Fig.~\ref{fig:SNR_error}.

As shown in Fig.~\ref{fig:SNR_error}, the relative uncertainties of all parameters generally decrease with increasing SNR.  
For the intrinsic SMBH parameters, the constraint on $m_{3}$ is significantly better than that on $\chi_{3}$.  
When the SNR is very low, the relative uncertainty of $\chi_{3}$ can even exceed 1, indicating that the spin of the SMBH cannot be effectively constrained by the GW signal alone.  
For future observations with an SNR threshold of 8, the median relative uncertainties of $m_{3}$ and $\chi_{3}$ are $1.8\times10^{-5}$ and $0.02$, respectively, which demonstrates a strong capability of GW observations to constrain SMBH properties. 

When the SNR exceeds 8, the median relative uncertainties reach $\sim 10^{-5}$ for $a_{\mathrm{out}}$ and $e_{\mathrm{out}}$, indicating extremely precise measurements.  
In contrast, the angular parameters $\zeta$ and $\epsilon$ are less well constrained, and their results are very similar to each other.  
Like $\chi_{3}$, these parameters are almost unmeasurable at very low SNR values, but once the SNR exceeds 8, their relative uncertainties reach a few percent.  
Since $\zeta$ and $\epsilon$ can describe the orientation of the SMBH spin, these results indicate that the spin direction can be constrained with reasonable accuracy through GW observations.

In terms of distance measurement, the constraint on $D_{L}$ is relatively weak.  
Only when the SNR exceeds 3 does its relative uncertainty drop below 1, indicating that $D_{L}$ becomes marginally measurable.  
Even at $\mathrm{SNR}>8$, the median relative uncertainty of $D_{L}$ is about 0.19.  
This limitation arises from the well-known degeneracy between $D_{L}$ and the inclination angle $\iota$ in GW observations~\cite{GW170817}.  
In contrast, the sky localization benefits significantly from the long observation time.  
When the SNR exceeds 8, the median sky localization accuracy reaches $3.9\times10^{-5}\, \mathrm{deg^2}$, allowing the source to be confined within a very small region of the sky. 

In summary, the results presented in this section demonstrate that space-based GW observations, particularly those with high SNR, can provide precise measurements of the SMBH mass and spin, as well as the outer orbital parameters of the hierarchical triple system.  
The luminosity distance remains the most weakly constrained parameter, while the angular parameters related to the SMBH spin orientation can still be constrained to a few percent accuracy.  
Overall, these findings indicate that long-duration space-based observations of stellar-mass BBHs orbiting SMBHs will offer a powerful means to probe the properties of central supermassive black holes with unprecedented precision.

\subsection{Application to M87*-like system}
In this section, we focus on SMBHs that have already been observed through EM means.  
Among them, the two most extensively studied cases are Sagittarius~A* at the center of the Milky Way and M87* at the core of the galaxy M87.  
These two SMBHs are also the only ones that have been directly imaged by the Event Horizon Telescope (EHT)~\cite{EHT_M87,EHT_SgrA}.  
Considering the typical population distribution of stellar-mass BBHs and the suitability for space-based GW detection, we select M87*-like system as a representative example.  
We investigate how GW observations can constrain its physical parameters and assess the potential improvement such measurements could provide to existing EM observations.

Such a configuration should be regarded as an idealized scenario. 
A simple order-of-magnitude estimate based on the local BBH merger-rate density ($R \sim 28.1\,\mathrm{Gpc^{-3}\,yr^{-1}}$) indicates that the total number of stellar-mass BBHs within the distance range corresponding to M87 ($\sim 15$--$18\,\mathrm{Mpc}$) and with coalescence time $\tau_c \in (0,10^4)\,\mathrm{yr}$ is only $N \sim 2.9$, and the number detectable by LISA would be even smaller.
Related studies of GW signals from triple systems in M87* can be found in Ref.~\cite{M87_GW}. 
This suggests that the system considered here serves as a benchmark configuration rather than a typical astrophysical expectation.

\begin{figure}[ht]
    \begin{minipage}{\columnwidth}
        \centering
        \includegraphics[width=0.95\textwidth,
        trim=0 0 0 0,clip]{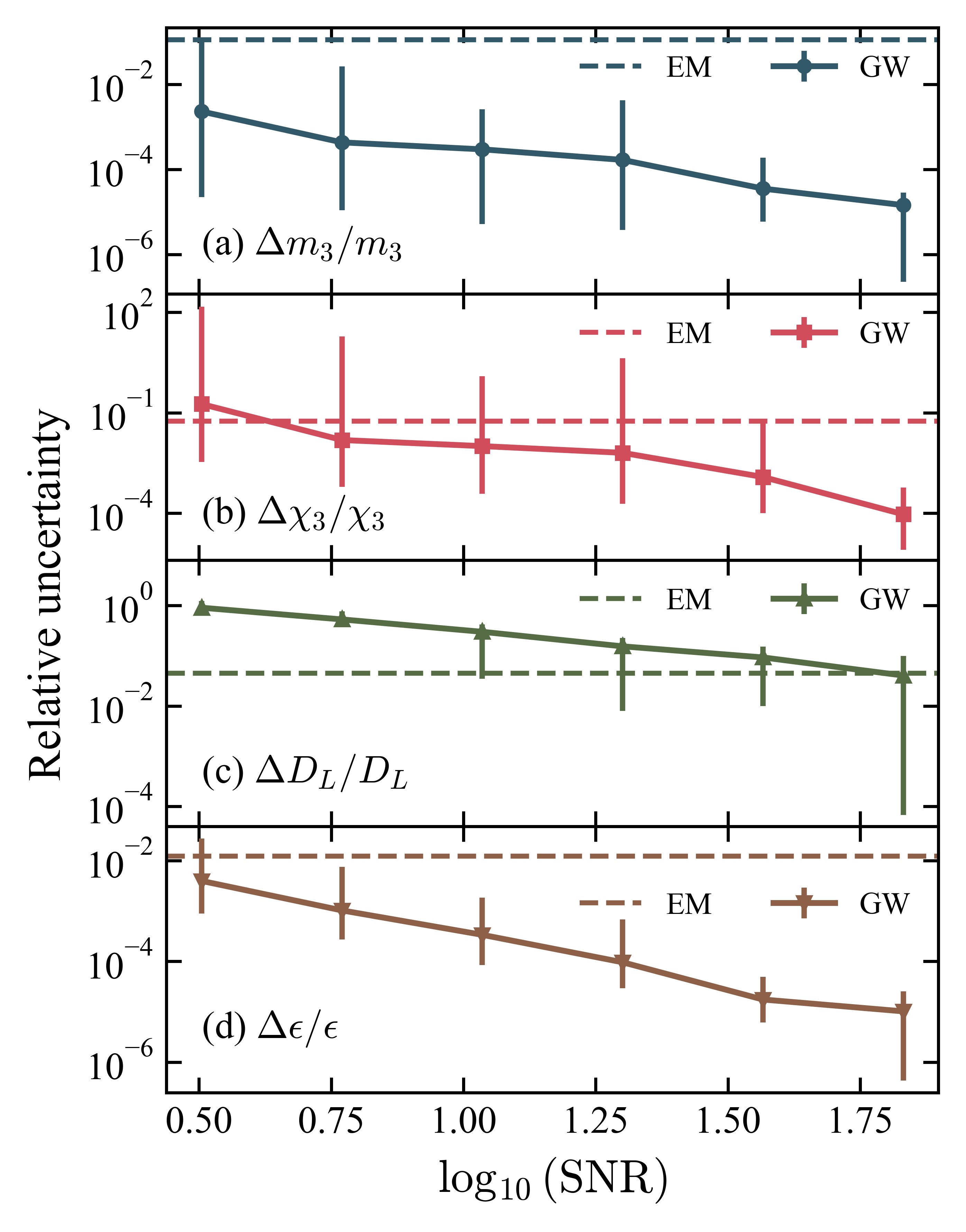}
        \caption{Relative uncertainties of the M87*-like parameters with varying SNR. Panels (a)–(d) show the results for $m_{3}$, $\chi_{3}$, $D_{L}$, and $\epsilon$, respectively. The plot elements are the same as those in Fig.~\ref{fig:SNR_error}. The dashed lines indicate the relative parameter uncertainties from the EM observations.
}\label{fig:M87_error}
    \end{minipage}
\end{figure}

We apply the same method to simulate GW observations, assuming the presence of a stellar-mass BBH continuously emitting GWs in the vicinity of M87*-like. 
The parameters of the SMBH are taken from the EHT observations, with  
$m_{3} = (6.5 \pm 0.72)\times10^{9}\,\mathrm{M_{\odot}}$,  
$D_{L} = 16.8^{+0.8}_{-0.7}\,\mathrm{Mpc}$,  
$\chi_{3} = 0.9 \pm 0.05$,  
and $\epsilon = 163^{\circ} \pm 2^{\circ}$~\cite{M87_mass,M87_spin}.
All other parameters follow the distributions listed in Table~\ref{tab:parameters}, while the BBH is assumed to be an equal-mass system with $m= 100\,\mathrm{M_{\odot}}$.
We simulate 500 hierarchical triple systems that satisfy the stability criterion to model the constraints on M87*-like parameters through GW observations.  
The results are shown in Fig.~\ref{fig:M87_error}.

GW observations provide much tighter constraints on the mass of M87*-like compared with EM measurements, with the relative uncertainties being several orders of magnitude smaller at all SNR levels.  
The constraint on $\chi_{3}$ is also quite good.  
Except for very low SNR cases where the GW precision is slightly worse, GW results generally outperform EM observations.  
For $D_{L}$, the situation is similar to that discussed in the previous section, where the constraints remain relatively weak.  
Although the relative uncertainties of $D_{L}$ are below 1, they are still larger than those obtained from EM data, even for high SNR.  
For the inclination angle $\epsilon$, GW observations yield consistently better results than EM measurements.  
Overall, except for $D_{L}$, the GW constraints on all parameters of M87*-like are superior to those from EM observations.

Furthermore, we investigate how GW observations can complement EM measurements, forming a joint multimessenger observation. 
The EM information refers to the observed properties of the SMBH and its host system, rather than to any electromagnetic counterpart of the stellar-mass BBH.
EM observations can constrain the source sky position with extremely high precision; for example, the EHT achieves an angular resolution of the order of microarcseconds~\cite{EHT_M87}.  
Therefore, we assume that the source position is well determined and treat it as fixed in our analysis.  
In the FIM calculation, this is implemented by removing the corresponding rows and columns associated with $\Phi$ and $\Theta$, which effectively reduces the parameter uncertainties~\cite{Wu_GB,FIM_ND}.  
The combined uncertainty from joint GW and EM observations is then estimated by summing the inverse variances, expressed as $1/\Delta \xi_{\mathrm{GW+EM}}^{2} = 1/\Delta \xi_{\mathrm{GW}}^{2} + 1/\Delta \xi_{\mathrm{EM}}^{2}$.  
This approach provides an approximate quantification of how GW observations can enhance EM measurements.  
The results are summarized in Table~\ref{tab:GW_EM}.

\begin{table}[ht]
\centering
\renewcommand{\arraystretch}{1.5}
\caption{Relative parameter uncertainties for EM-only, GW-only, and combined GW+EM observations.  
The GW results correspond to cases with $\mathrm{SNR}>8$.}\label{tab:GW_EM}
\begin{tabular*}{\columnwidth}{@{\extracolsep{\fill}}cccc@{}}
\hline
 $\Delta \xi/\xi$ & EM & GW & GW+EM \\
\hline
$m_3$ & $1.12 \times 10^{-1}$ & $1.79 \times 10^{-4}$ & $1.51 \times 10^{-4}$\\
$\chi_3$ & $5.56 \times 10^{-2}$ & $5.49 \times 10^{-3}$ & $5.15 \times 10^{-3}$\\
$D_L$ & $4.46 \times 10^{-2}$ & $1.87 \times 10^{-1}$ & $4.34 \times 10^{-2}$\\
$\epsilon$ & $1.23 \times 10^{-2}$ & $1.24 \times 10^{-4}$ & $8.86 \times 10^{-5}$\\
\hline
\end{tabular*}
\end{table}

It can be seen that, except for $D_{L}$, the relative uncertainties obtained from GW observations are much smaller than those from EM measurements.  
In the case of joint multimessenger observations, these constraints can be further improved.  
Our calculations show that the relative uncertainty of $m_{3}$ can be reduced to about 0.1\% of the EM value, representing an improvement of three orders of magnitude.  
The uncertainties of $\chi_{3}$ and $\epsilon$ also decrease to approximately 9.3\% and 0.7\% of the EM results, respectively, significantly enhancing the measurement precision of M87*-like.  
In contrast, the improvement for $D_{L}$ is only about 3\%, which is much smaller than that achieved for the other parameters.

In summary, if a stellar-mass BBH exists around M87*-like and forms a hierarchical triple system, it has the potential to be detected by future space-based GW detectors.  
Such a detection would enable a much more precise measurement of the SMBH parameters compared with current EM observations.  
This improvement would not only advance our understanding of GW sources and their dynamics but also provide valuable support for EM studies.  
By combining the two observation channels, tighter constraints on the SMBH mass, spin, and orientation could be achieved, offering critical insights into the physical processes in galactic nuclei and helping to refine theoretical models of galaxy formation and the coevolution of galaxies and their central black holes.

\section{Conclusion}\label{Conclusion}
In this paper, we investigate how the modulation of GWs emitted by a stellar-mass BBH orbiting an SMBH can be used to constrain the SMBH parameters through observations by the space-based detector LISA.  
We focus on dynamically stable hierarchical triple systems and numerically integrate the geodesic equations in the Kerr metric to obtain the non-equatorial, eccentric orbital trajectories and corresponding velocity components.  
The intrinsic waveform of the BBH is modeled using the 3.5PN+SO+SS approximation, which describes the inspiral phase of a stationary binary.  
By applying rotation and moving-source transformations, we compute the modulated GW waveform from a moving BBH.  
Using the LISA orbital configuration and noise model, we calculate the detector responses in the TDI 2.0 framework and evaluate the resulting signals for up to 4 years.  
The parameter estimation is performed through the calculation of the SNR and the FIM, which provide the parameter uncertainties and thus quantify LISA’s capability to constrain the SMBH properties.  
Within the parameter space considered, we simulate two thousand hierarchical triple systems and analyze the effects of different parameters on the measurement precision.  
In addition, we apply our method to the well-studied SMBH M87*-like to assess the potential constraints achievable through future GW observations.

Our results demonstrate the great potential of future space-based GW detectors in constraining SMBH parameters.  
Among the outer orbital parameters of the SMBH–BBH systems, the outer semimajor axis has the strongest impact on the parameter estimation accuracy, while the SMBH spin and outer eccentricity mainly affect their own uncertainties.  
Further analysis shows that at high SNR, LISA can constrain the SMBH spin magnitude and orientation to within a few percent, while the SMBH mass, outer semimajor axis, and eccentricity can be measured with relative uncertainties on the order of $10^{-5}$.  
The sky localization accuracy can also reach $\sim10^{-5}\,\mathrm{deg^{2}}$, although the constraint on the luminosity distance remains relatively weak.  
When applied to M87*-like, GW observations can provide constraints several orders of magnitude tighter than current EM measurements, particularly for the SMBH mass and spin.  
In a multimessenger context, the combination of GW and EM data can further improve measurement precision, offering a powerful approach to studying SMBHs and the stellar-mass BBHs orbiting them.  
Overall, this work highlights the capability of space-based GW astronomy to probe SMBH properties with unprecedented accuracy and to complement EM observations in understanding the formation and evolution of galactic nuclei.

In future research, we plan to further extend this work.  
First, we will incorporate more complete physical effects on the BBH, such as de Sitter precession, to obtain a more comprehensive description of the waveform.  
We also aim to include internal effects of the BBH, such as inner orbital eccentricity and spin precession, to better capture the intrinsic dynamics of the inner binary.
Moreover, implementing more realistic detector configurations, data analysis pipelines, and Bayesian inference methods will help translate these theoretical predictions into observational forecasts.  
Finally, extending this framework to other planned space missions, such as Taiji and TianQin, will enable a comprehensive assessment of the combined sensitivity, providing deeper insight into the dynamics of hierarchical triple systems and the astrophysical population of SMBHs throughout the Universe.

\begin{acknowledgements}
This work was supported by the National Key Research and Development Program of China (Grant No. 2023YFC2206702), the National Natural Science Foundation of China (Grant Nos. 125B2102, 12575072 and 12547101), the Fundamental Research Funds for the Central Universities Project (Grant No. 2024IAIS-ZD009), and the Natural Science Foundation of Chongqing (Grant No. CSTB2023NSCQ-MSX0103).
\end{acknowledgements}
\begin{widetext}
\appendix
\section{Coordinate transformation}\label{App:Coordinate_transformation}
The SMBH and SSB coordinate systems illustrated in Fig.~\ref{fig:SMBH-BBH}(b) can be related through a sequence of Euler rotations followed by a rotation about the propagation direction $\hat{r}$.  
For a vector expressed in the SMBH coordinate system as $\mathbf{v}_{xyz} = [v_x, v_y, v_z]^T$, its corresponding components in the SSB coordinate system, $\mathbf{v}_{XYZ} = [v_X, v_Y, v_Z]^T$, can be obtained through the transformation  
\begin{equation}\label{eq:R}
    \mathbf{v}_{XYZ} = \mathbf{R}(\Theta, \Phi, \epsilon, \zeta)\,\mathbf{v}_{xyz},
\end{equation}
where $\mathbf{R}(\Theta, \Phi, \epsilon, \zeta)=\mathbf{R}_o(\Theta,\Phi,\epsilon)
      \, \mathbf{R}_s(\epsilon,\zeta)$ represents the combined rotation matrix defined by the Euler angles $(\Theta, \Phi, \epsilon)$ and the additional rotation $\zeta$ about $\hat{r}$.
The explicit forms of the rotation matrices $\mathbf{R}_s$ and $\mathbf{R}_o$ are given below.
\begin{equation}
\mathbf{R}_s(\epsilon,\zeta)=
\begin{bmatrix}
\cos\zeta & -\sin\zeta\cos\epsilon & \sin\zeta\sin\epsilon\\[4pt]
\sin\zeta\cos\epsilon & \cos\zeta+(1-\cos\zeta)\sin^2\epsilon & (1-\cos\zeta)\sin\epsilon\cos\epsilon\\[4pt]
-\sin\zeta\sin\epsilon & (1-\cos\zeta)\sin\epsilon\cos\epsilon & \cos\zeta+(1-\cos\zeta)\cos^2\epsilon
\end{bmatrix},
\end{equation}

\begin{equation}
\mathbf{R}_o(\Theta,\Phi,\epsilon)=
\begin{bmatrix}
-\sin\Phi & \cos\Phi\cos(\Theta+\epsilon) & -\cos\Phi\sin(\Theta+\epsilon)\\[4pt]
\cos\Phi & \sin\Phi\cos(\Theta+\epsilon) & -\sin\Phi\sin(\Theta+\epsilon)\\[4pt]
0 & -\sin(\Theta+\epsilon) & -\cos(\Theta+\epsilon)
\end{bmatrix}.
\end{equation}
The definitions of the involved angles are illustrated in Fig.~\ref{fig:SMBH-BBH}.
The inverse transformation from the SSB coordinate system to the SMBH coordinate system can be obtained simply by taking the inverse of Eq.~(\ref{eq:R}), that is, $\mathbf{v}_{xyz} = \mathbf{R}^{-1}\,\mathbf{v}_{XYZ}$.
\end{widetext}
\bibliography{references}
\end{document}